\documentclass{scrartcl}
\usepackage[affil-it]{authblk}
\usepackage{hyperref}
\hypersetup{colorlinks=true,urlcolor=blue}
\begin{document}
\title{Notes on the page proof and publication of ``Schallausbreitung in teilweise dissoziierten Gasen'' by A. Einstein}
\author{J. L. van Velsen}
\affil{Instituut-Lorentz, Universiteit Leiden, P.O. Box 9506\\
2300 RA Leiden, The Netherlands}
\date{November 2005}
\maketitle

Recently, the \href{http://www.lorentz.leidenuniv.nl/history/Einstein_archive}{page-proof}, with corrections, of Einstein's publication\footnote{A. Einstein, \textit{Schallausbreitung in teilweise dissoziierten Gasen}, Sitzungsberichte der Preu{\ss}ischen Akademie der Wissenschaften, Berlin, 8 April 1920, p.\ 380--385.} on the ``Propagation of Sound in Partly Dissociated Gases'', which appeared in 1920 
in the Proceedings of the Berlin Academy of Sciences, was found at the Instituut-Lorentz of
Leiden University.\footnote{Archived in the Leiden University Library, code number BPL 3545.} The last two pages of the proof are crossed out and replaced by a sheet in 
Einstein's handwriting. In these notes, I briefly address the following two questions:
\begin{itemize}
\item{Concerning the page proof, what is the difference between the contents of the crossed-out pages and the new page?}
\item{Concerning the publication, are Eqs.\ (18,21,22) correct?}
\end{itemize}

I start with the second question. The \href{http://articles.adsabs.harvard.edu/cgi-bin/get_file?pdfs/SPAW./1920/1920SPAW.......380E.pdf}{scanned publication} provided by the Max Planck Institute for the History of Science in Berlin as part of the {\sc echo} initiative (European Cultural Heritage Online), contains hand-made changes to Eqs.\ (18,21,22). After doing the calculations myself, I found these changes to be correct. 
The corrections affect the final result of the publication, the speed of sound $V$.

Regarding to the first question, the crossed-out pages deal with $V$ in the special case $RT \ll D$, while the new page contains a more
general result for $V$, not restricted to the regime $RT \ll D$. In addition, the crossed-out pages contain errors.
The speed of sound $V_{\rm new}$ on the new page and the speed of sound $V_{\rm old}$ on the crossed-out pages are given by
\begin{equation}
V^{2}_{\rm new}=\frac{p}{\rho}\left(1+\frac{x_{1}^{2}AB+R\overline{c}\omega^{2}}{x_{1}^{2}B^{2}+{\overline{c}}^{2}\omega^{2}}\right), \quad
V^{2}_{\rm old}=\frac{p}{\rho}\left(1+\frac{R\overline{c}\omega^{2}}{\left(x_{1}\eta_{1}\frac{D^{2}}{RT^{2}}\right)^{2}+\overline{c}\omega^{2}}\right),
\end{equation}
with
\begin{equation}
\overline{c}=\frac{C}{n_{1}+n_{2}},
\end{equation}
\begin{equation}
A=\left( \frac{2D}{T}-\overline{c}\right)\frac{n_{1}}{n_{1}+n_{2}}+R\left(1 \pm \frac{4n_{1}}{n_{2}}\right),
\end{equation}
\begin{equation}
B=\frac{D^{2}}{RT^{2}}\frac{n_{1}}{n_{1}+n_{2}}+\overline{c}\left(1 \pm \frac{4n_{1}}{n_{2}}\right).
\end{equation}
(The ``$\pm$'' refers to the error in Eq.\ (18): Einstein took the minus sign, it should be the plus sign.) 
Since both $V_{\rm old}$ and $V_{\rm new}$ originate from Eq.\ (18), we can compare $V_{\rm old}$ to $V_{\rm new}$ in the 
regime $RT \ll D$. The differences between the two are due to errors made in the calculation of $V_{\rm old}$ on the crossed-out pages:
\begin{itemize}
\item{A factor $\eta_{1}+\eta_{2}$ has not been properly devided out in the transition from Eq.\ (18a) to the next equation. This accounts for
the difference between $x_{1}^{2}B^{2}$ and $\left(x_{1}\eta_{1}\frac{D^{2}}{RT^{2}}\right)^{2}$ in denominators in $V_{\rm new}$
and $V_{\rm old}$, respectively.}
\item{The term $\overline{c}\omega^{2}$ in the denominator in $V_{\rm old}$ should be ${\overline{c}}^{2}\omega^{2}$. In the limit $\omega \rightarrow \infty$, 
$V_{\rm old}$ takes the wrong value.} 
\item{The contribution $x^{2}_{1}AB$ to the numerator in $V_{\rm old}$ is ignored. This can be traced back to the transition from the intermediate equation
between Eqs.\ (18a) and (19) to Eq.\ (19).} 
\end{itemize}

\noindent {\rm CONCLUSIONS:}

\begin{itemize}
\item{Concerning the page proof, what is the difference between the contents of the crossed-out pages and the new page? {\em Answer:}
\vspace{-1mm}
\begin{itemize}
\item{The new page is a generalization of the crossed-out pages, which deal with a special regime.}
\item{The crossed-out pages contain errors.}
\end{itemize}}
\item{Concerning the publication, are Eqs.\ (18,21,22) correct? {\em Answer:} 
\vspace{-1mm}
\begin{itemize}
\item{Eqs.\ (18,21,22) are not correct. However, the scanned version with the hand-made changes is correct.}
\end{itemize}}
\end{itemize}
\end{document}